\documentclass[aritcle,aps,epsf,amsfonts,amssymb,amsmath,nofootinbib]{revtex4} 

\usepackage{amsmath,amssymb,amsfonts,amsthm}
\usepackage{graphicx}

\usepackage{subeqnarray}

\newcommand{\be}{\begin{equation}}\newcommand{\ee}{\end{equation}}
\newcommand{\bea}{\begin{eqnarray} }\newcommand{\eea}{\end{eqnarray}}
\newcommand{\beaa}{\begin{eqnarray} }\newcommand{\eeaa}{\end{eqnarray}}
\newcommand{\bsa}{\begin{subeqnarray}}\newcommand{\esa}{\end{subeqnarray}}
\newcommand{\ba}{\begin{array}}\newcommand{\ea}{\end{array}}
\newcommand{\bit}{\begin{itemize}}\newcommand{\eit}{\end{itemize}}
\newcommand{\ben}{\begin{enumerate}}\newcommand{\een}{\end{enumerate}}

\def\lab{\label}

\def\rar{\rightarrow}

\def\al{\alpha}\def\ga{\gamma}\def\Ga{\Gamma}
\def\de{\delta}
\def\ka{\kappa}
\def\om{\omega}
\def\Om{\Omega}%

\def\1{{_{1}}}\def\2{{_{2}}}

\newcommand{\half}{\frac{1}{2}}

\def\lsim{\hbox{ \raise.35ex\rlap{$<$}\lower.6ex\hbox{$\sim$}\ }}
\def\gsim{\hbox{ \raise.35ex\rlap{$>$}\lower.6ex\hbox{$\sim$}\ }}

\def\lab{\label}

\def\rar{\rightarrow}

\def\al{\alpha}

\def\ga{\gamma}
\def\Ga{\Gamma}
\def\de{\delta}

\def\ka{\kappa}

\def\om{\omega}
\def\Om{\Omega}
\begin{document}
\title{Fractals, coherent states and self-similarity induced noncommutative geometry
}

\author{Giuseppe Vitiello\footnote{vitiello@sa.infn.it ~~--~~ www.sa.infn.it/giuseppe.vitiello/}}
\affiliation{Dipartimento di Fisica ``E. R. Caianiello'' and Istituto Nazionale di Fisica Nucleare\\ Universit\'a di Salerno, I-84084 Fisciano (Salerno), Italy}

\begin{abstract}
The self-similarity properties of fractals are studied in the framework of the theory of entire analytical functions and the $q$-deformed algebra of coherent states. Self-similar structures are related to dissipation and to noncommutative geometry in the plane. The examples of the Koch curve and logarithmic spiral are considered in detail. It is suggested that the dynamical formation of fractals originates from the coherent boson condensation induced by the generators of the squeezed coherent states, whose (fractal) geometrical properties thus become manifest. The macroscopic nature of fractals appears to emerge from microscopic coherent local deformation processes.

\vspace{5mm}
\end{abstract}

\maketitle

\section{Introduction}\label{sec1}

Recently, self-similarity properties of
deterministic fractals have been studied in
the framework of the theory of entire analytical functions and their functional realization in terms of the $q$-deformed
algebra of squeezed coherent states has been discussed~\cite{NewMat2008}. These studies are further pursued in the present paper where  self-similar  structures are related to the notion
of dissipative quantum interference phase and noncommutative geometry. The results presented below are  framed in the research line which has shown that {\it macroscopic quantum systems}
or ``extended objects'', such as kinks, vortices, monopoles, crystal dislocations, domain walls, and other so-called ``defects'' in condensed matter physics are generated  by coherent quantum condensation processes at the microscopic level~\cite{Umezawa:1982nv,DifettiBook,Bunkov}. The theoretical description so obtained has been quite successful in explaining many experimental observations, e.g., in superconductors, crystals, ferromagnets, etc.~\cite{Bunkov}. This motivates the present attempt to frame in such a dynamical description also self-similar structures, such as fractals (to the extent in which fractals are considered under the point of view of self-similarity, considering that in some sense self-similarity is  the {\it most important property} of fractals (p. 150 in Ref.~\cite{Peitgen})). Such an attempt is also in agreement with and supported by recent observations~\cite{CrystalFractals} showing that
when a crystal is bent (submitted to deforming stress actions), the so produced crystal defects in the lattice (dislocations)  form, at low temperature, self-similar fractal patterns, which thus appear as the result of non-homogeneous coherent phonon (boson) condensation~\cite{Umezawa:1982nv,DifettiBook} and provide an example of ``emergence of fractal dislocation structures''~\cite{CrystalFractals}
in nonequilibrium (dissipative) systems.
On the other hand, the possibility of relating some properties of  fractals to coherent states and dissipative systems is {\it per se} of great theoretical and practical interest due to the ubiquitous presence of fractals in nature~\cite{Bunde,Peitgen}  and the relevance of coherent states and dissipative systems in a wide range of physical phenomena.

In the following, the self-similarity properties of the well known examples of the Koch curve (Fig.~\ref{fig1}) and the  logarithmic spiral (Fig.~\ref{fig2}) are considered.
Their relation to squeezed coherent states, dissipation and noncommutative geometry  is presented in Section II, III and IV, respectively. Section V is devoted to conclusions. The measure of lengths in fractals, the Hausdorff measure, the fractal ``mass'', random fractals, and other fractal properties are not discussed in this paper. The discussion is limited to the self-similarity properties of deterministic fractals (which are generated iteratively according to a prescribed recipe).

\section{Self-similarity and coherent states}\label{sec2}

\begin{figure}
\centering \resizebox{6cm}{!}{\includegraphics{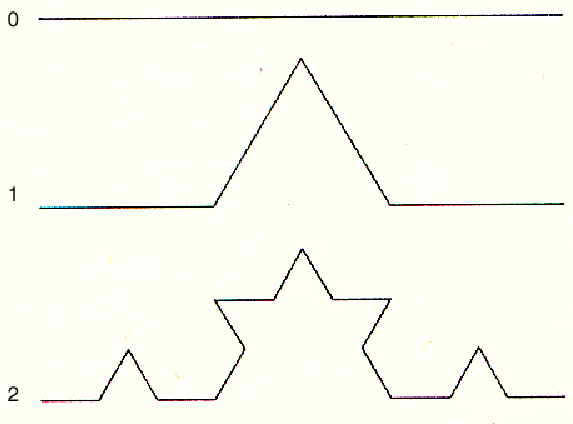}}
\centering \resizebox{6cm}{!}{\includegraphics{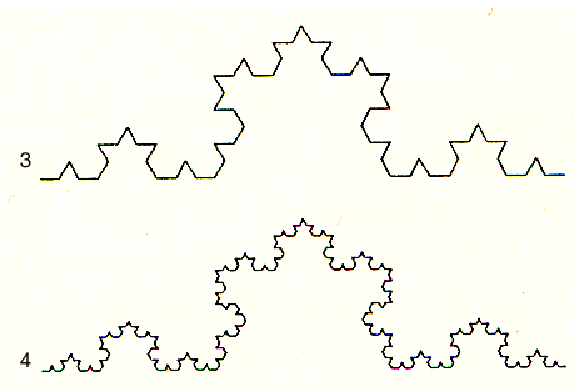}}
\caption{\small \noindent The first five stages of Koch curve.}
\label{fig1}
\end{figure}

Some of the results relating the self-similarity properties of Koch curve to the $q$-deformed coherent states are summarized briefly in the following (the conclusions can be extended to other known examples, such as the Sierpinski gasket and carpet, the Cantor set, etc.).
The construction of the Koch curve can be done by adopting the standard procedure~\cite{Bunde} (see also \cite{NewMat2008,QI}) leading to the {\it fractal dimension} $d$, also called the
{\it self-similarity dimension} \cite{Peitgen},
\be \lab{30} d = \frac{\ln 4}{\ln 3} \approx 1.2619~. \ee
Let the $n$-th step or stage of the Koch curve construction (see Fig.~\ref{fig1}) be denoted by $u_{n,q}(\alpha)$, with $\alpha = 4$ and $q = 1/3^d$. Setting the starting stage $u_{0} = 1$,
one has
\be \lab{31} u_{n,q}(\alpha)  = (q \, \alpha)^{n} = 1  , ~~ \quad {\rm for ~any} ~n, \ee
from which Eq.~(\ref{30}) is obtained. It has to be stressed that self-similarity is properly defined only in the $n \rar \infty$ limit  (self-similarity does not
hold when considering only a finite number $n$ of iterations).

By considering in full generality the complex
$\al$-plane, and putting $q = e^{-d \, \theta}$, Eq.~(\ref{30}) is written as $d \, \theta = \ln \, \al$ and the functions $u_{n,q}(\alpha)$
are, apart the
normalization factor $1/\sqrt{n!}$,
nothing but the restriction to real
$\al$ of the functions
\be \lab{(a2.8)} u_{n,q}(\alpha) = {(q\, \al)^n \over \sqrt{n!}} ~,~\quad
\quad \quad~~ n \in \mathcal{N}_+ ~, \quad \al \in
{\bf  \mathbb{C}} ~,
\ee
which form a basis in the space ${\cal F}$ of the entire analytic functions,
orthonormal under the gaussian measure $d\mu(\al)=$
$\displaystyle{({1/{\pi})} e^{- |\al|^2} d\al d{\bar \al}}$.
The study of the fractal
properties may thus be carried on in the space ${\cal F}$ of the
entire analytic functions, by restricting, at the end, the
conclusions to real $\al$, $\al \rar {\it Re}(\al)$~\cite{NewMat2008,QI}.
The connection between fractal self-similarity properties and coherent states is then readily established since one realizes that
the space ${\cal F}$ is
the vector space which provides the so-called Fock-Bargmann
representation (FBR) of the Weyl--Heisenberg
algebra \cite{Perelomov:1986tf} and is an useful frame to
describe  the (Glauber) coherent states. The ``$q$-deformed'' algebraic structure of which the space
${\cal F}$ provides a representation is obtained by introducing
the finite difference operator ${\cal D}_q$,
also called the $q$-derivative operator \cite{13}:
\be \lab{(2.12)} {\cal D}_q f(\al) = {{f(q \al) - f(\al)}\over
{(q-1) \al}} ~, \ee
with ~$f(\al) \in {\cal F}\; ,\; q = e^\zeta \; ,\; \zeta \in {{\bf
\mathbb{C}}}$ . ${\cal D}_q$ reduces to the standard derivative for $q \to 1$ ($\zeta \to 0$). By applying $q^{N}$ to the coherent state $|\al \rangle$ ($a |\al \rangle = \al |\al \rangle$, with $a$ the annihilator operator), one finds that
\be \lab{(a2.21)}  q^{N} |\al \rangle = |q \al \rangle =
\exp\biggl(-{{|q\alpha|^2}\over 2}\biggr) \sum_{n=0}^\infty \frac{(q
\alpha)^{n}}{\sqrt{n!}}~ |n\rangle \ee
where $N \equiv \al\, {d/ d\al} $.
The $n$-th iteration stage of the fractal now can be ``seen'' by
applying $(a)^{n}$ to $|q \al \rangle$ and restricting to real $q
\al$
\be \lab{(ab2.24)}  \langle q \al | (a)^{n} |q \al \rangle =  (q
\alpha)^{n} = u_{n,q} (\al), ~~ \qquad q \al \rar {\it Re} (q \al). \ee
The operator $(a)^{n}$ thus acts as a ``magnifying'' lens
\cite{Bunde,NewMat2008,QI}.
Thus the  fractal $n$-th stage of iteration, with $n =
0,1,2,..,\infty$, is represented, in a one-to-one correspondence, by
the $n$-th term in the coherent state series
Eq.~(\ref{(a2.21)}).
The operator $q^{N}$
is called {\it the fractal
operator}~\cite{NewMat2008,QI}. Note that $|q \al \rangle$ is actually a squeezed coherent state\cite{CeleghDeMart:1995,Yuen:1976vy}. $\zeta = \ln q $ is called the
squeezing parameter and $q^N$
acts in ~${\cal F}$
as the squeezing operator.

The conclusion is that the self-similarity properties of the Koch curve (and other fractals) can be described in terms of the coherent state squeezing transformation.
Their functional realization in terms of the two mode $SU(1,1)$ representation is discussed in Section III.

These results can be extended also to the logarithmic spiral. Its
defining equation in polar coordinates $(r, \theta)$ is~\cite{Peitgen,Andronov}:
\be \lab{losp2.1} r = r_0 \, e^{d \, \theta} ~,
\ee
with  $r_0$ and  $d$ arbitrary real constants and $r_{0} > 0$, whose representation is the straight line of slope $d$ in a log-log plot with abscissa $\theta = \ln e^{\theta}$:
\be \lab{losp2.2} d \, \theta = \ln \frac{r}{r_0}~.
\ee
The constancy of the angular coefficient $\tan^{-1} d$ signals the self-similarity property of the logarithmic spiral: rescaling $\theta \rar n \, \theta$ affects $r/r_0$ by the power $(r/r_0)^n$.
Thus, we may proceed again like in the Koch curve case and show the relation to squeezed coherent states.
The result also holds for the specific form of the
logarithmic spiral called the {\it golden spiral} (see Appendix A).  The  golden spiral  and its relation to the Fibonacci progression is of great interest since the Fibonacci progression appears in many phenomena, ranging from botany to physiological and functional properties in living systems,  as the ``natural'' expression in which they manifest themselves. Even in linguistics, some phenomena have shown Fibonacci progressions~\cite{Piattelli}.

\begin{figure}
\centering \resizebox{7cm}{!}{\includegraphics{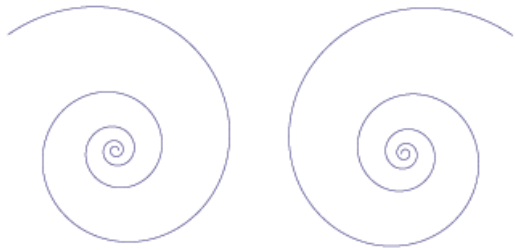}}
\caption{\small \noindent The anti-clockwise and the clockwise logarithmic spiral.}
\label{fig2}
\end{figure}

As customary, let in Eq.~(\ref{losp2.1}) the anti-clockwise angles $\theta$'s be taken to be positive.  The  anti-clockwise versus (left-handed) spiral has $q \equiv e^{d \, \theta} > 1$; the clockwise versus (right-handed) spiral has $q < 1$ (see Fig.~\ref{fig2}); sometimes they are called the {\it direct} and  {\it indirect} spiral, respectively.

In the next Section we consider the parametric equations of the spiral:
\bsa \lab{losp2.6} x &=& r (\theta) \, \cos \theta = r_0 \,e^{d \, \theta}\, \cos \theta ~, \\
  y &=& r (\theta) \, \sin \theta = r_0 \,e^{d \, \theta}\, \sin \theta  ~.
\esa

\section{Self-similarity, dissipation and squeezed coherent states}

According to Eqs.~(\ref{losp2.6}), the point on the logarithmic spiral in the complex $z$-plane is given by
\be \lab{losp2.7} z =  x  + i\, y =  r_0 \, e^{d \, \theta} \, e^{i \, \theta} ~,
\ee
The point $z$ is fully specified only when the sign of $d \, \theta$ is assigned. The factor $q = e^{ d \, \theta}$ may denote indeed one of the two components of the (hyperbolic) basis $\{e^{-  \,  d\, \theta}, \, e^{+  \, d\, \theta} \}$.
Due to the completeness of the basis,  both the factors  $e^{\pm \, d \, \theta}$ must be considered. It is indeed interesting that in nature in many instances the direct ($q > 1$) and the indirect ($q < 1$) spirals are both realized in the same system (perhaps the most well known systems where this happens are found in phyllotaxis studies).
The points $z_1$ and $z_2$ are then considered:
\be \lab{losp2.8} z_1 =  r_0 \, e^{- \, d \, \theta} \, e^{- \,i \, \theta}~, \qquad z_2 =   r_0 \, e^{+ \, d \, \theta} \, e^{+ \, i \, \theta}~,
\ee
where for convenience (see below)  opposite signs for the imaginary exponent $i \, \theta$ have been chosen. By using the parametrization  $\theta = \theta(t)$,  $z_1$ and $z_2$ are easily shown to solve the equations
\bsa \lab{losp2.10} m \, \ddot{z}_1 \, + \, \ga \, \dot{z}_1 \, + \, \kappa \, z_1  &=& 0 ~,\\ \lab{losp2.10b}
m \, \ddot{z}_2  \, - \, \ga \, \dot{z}_2 \, + \, \kappa \, z_2  &=& 0 ~,
\esa
respectively, provided the relation
\be \lab{losp2.11} \theta(t) =   \frac{\ga}{2 \, m \, d} \, t =  \frac{\Ga}{d} \, t
\ee
holds (up to an arbitrary additive constant $c$ which we set equal to zero). We see that $\theta (T) = 2 \, \pi$ at  $T = 2\, \pi \, d/ \Ga$.
In Eqs.~(\ref{losp2.10}) ``dot" denotes derivative with respect to $t$;   $m$, $\ga$ and $\kappa$ are positive real constants. The notation $\Ga \equiv {\ga\over 2 m}$ also will  be used. One can also show that $\rho_\pm (t) \equiv r_0 \, e^{\pm \,i\, \theta (t)}$ satisfy the equation
\be \lab{losp2.12} m\,\ddot{\rho}_\pm  \, + \, K \, \rho_\pm = 0 ~,
\ee
with
\be \lab{losp2.13} K = m {\Om}^2 ~, \qquad \Om^2 = \frac{1}{m}
(\ka-\frac{\ga^2}{4m}) =  \frac{\Ga^2}{ d^2}~,~\qquad \mbox{with}~~ \ka >\frac{\ga^2}{4m}~.
\ee
In conclusion, the parametric expressions $z_1 (t)$ and $z_2 (t)$ for the logarithmic spiral are $z_1 (t) = r_0 \,  \, e^{- \,i \, \Om \, t}\, e^{- \Ga t }$ and $z_2  (t) = r_0 \,  \, e^{+ \,i \, \Om\,t}\, e^{ + \Ga \, t }$, solutions of Eqs.~(\ref{losp2.10}a) and (\ref{losp2.10}b). At $t = m \, T$, $z_1 = r_0 \, (e^{ -\,2 \, \pi \,  d })^m $, $z_2 = r_0 \, (e^{ 2 \, \pi \, d })^m $, with the integer $m = 1,\, 2,\, 3$...

These considerations and Eqs.~(\ref{losp2.10}) suggest to us that one can interpret the parameter $t$ as the time parameter. As a result, we have that the time-evolution of the system of direct and indirect spirals is described by the system of equations (\ref{losp2.10}a) and (\ref{losp2.10}b) for the damped and amplified harmonic oscillator. The spiral ``angular velocity" is given by $|\, d \,\theta/dt \,|  = |\, \Ga/d \,|$.
Note that the oscillator $z_1$ is an {\it open} (non-hamiltonian) system and it is well known~\cite{Celeghini:1992yv} that in order to set up the canonical formalism one needs to {\it double} the degrees of freedom of the system by introducing its time-reversed image $z_2$: we thus see that the physical meaning of the mentioned mathematical necessity to consider both the elements of the basis $\{e^{-\, d\, \theta}, \, e^{+ \, d \, \theta} \}$ is that one needs to
consider the {\it closed} system $(z_1 ,\, z_2)$~\cite{Celeghini:1992yv} since then we are able to set up the canonical formalism. The  closed system Lagrangian is then given by
\be \lab{losp2.14}
L = m \dot z_1 \dot z_2 + \half \gamma ( z_1 \dot z_2 - \dot z_1 z_2 ) - \kappa z_1 z_2~,
\ee
from which,
Eqs.~(\ref{losp2.10}a) and (\ref{losp2.10}b)) are both derived
by varying $L$ with respect to $z_2$ and $z_1$, respectively.
The canonical momenta
are given by
\be \lab{losp2.14p}
p_{z_1} = \frac{\partial L}{\partial {\dot z}_1 } = m {{\dot z}_2} - \frac{1}{2}
\gamma {z_2}~, \qquad p_{z_2} = \frac{\partial L}{\partial {\dot z}_2 }
= m {{\dot z}_1} + \frac{1}{2} \gamma {z_1}~.
\ee

It is worth stressing that without closing the system $z_1$ with the system $z_2$ there would be no possibility to define the momenta $p_{z_i}$, $i = 1,\,2$, since  the same Lagrangian $L$ would not exist.
The time evolution of the ``two copies"
$(z_1 ,\, z_2)$ of the $z$-coordinate
can be viewed as the forward in time path
and the backward in time path in the phase space $\{z, p_z \}$,
respectively. We thus arrive  at a crucial point in our discussion: it is known indeed that
as far as  $z_1 (t) \neq z_2 (t)$ the system exhibits quantum behavior and quantum interference takes place~\cite{Blasone:1998xt,Srivastava:1995yf,Links,DifettiBook}. By following
Schwinger~\cite{Schwinger}, this can be explicitly proven, for example, by considering the double slit experiment in quantum mechanics~\cite{Blasone:1998xt,Srivastava:1995yf,Links,DifettiBook}.
In the quantum mechanical formalism of the Wigner function and density matrix it is required to
consider doubling the system coordinates, from one coordinate $z(t)$
describing motion in time to two coordinates, $z_1(t)$ going forward in time and
$z_2 (t)$ going backward in time~\cite{Blasone:1998xt,Srivastava:1995yf}. In order to have quantum interference, the forward in time action $A(z - z_1 , t)$ must be different from the backward in time action $A(z - z_2 , t)$.
The classical behavior of the system
is obtained if the two paths can be identified, i.e. $ z_1 (t) \approx z_2 (t) \approx z_{classical} (t)$. When  the forward in time and backward in time motions are (at the same time)
unequal $z_1 (t) \neq z_2 (t)$, then the system behaves in a quantum mechanical
fashion. Of course, when $z$ is actually measured there is only one classical $z_{classical}$. This picture also agrees with 't Hooft conjecture, which states that, provided some specific energy conditions are met and
some constraints are imposed, classical, deterministic systems presenting loss of information (dissipation) might behave according to a quantum evolution~\cite{'tHooft:1999gk,Blasone:2000ew}. In the present case, this means that
the logarithmic  spiral and its time-reversed double (the anti-clockwise spiral and the clockwise spiral) manifest themselves as {\it macroscopic quantum systems}.

The canonical
commutators $[\, z_1 ,\, p_{z_1} \, ] = i\, \hbar = [\, z_2 ,\, p_{z_2} \,] , ~
[\,z_1 ,\, z_2 \,] = 0 = [\, p_{z_1} , p_{z_2} \, ]$, and the corresponding sets of
annihilation and creation operators are introduced in a standard fashion:
\bea \lab{losp2.19}
a &\equiv& \left ({1\over{2 \hbar \Omega}} \right )^{1\over{2}} \left (
{{p_{z_1}}\over{\sqrt{m}}} - i \sqrt{m} \Omega z_1 \right ) ~ ; \qquad
a^{\dagger} \equiv \left ({1\over{2 \hbar \Omega}} \right )^{1\over{2}} \left (
{{p_{z_1}}\over{\sqrt{m}}} + i \sqrt{m} \Omega z_1\right ) ~;\\
b &\equiv& \left ({1\over{2 \hbar \Omega}} \right )^{1\over{2}} \left (
{{p_{z_2}}\over{\sqrt{m}}} - i \sqrt{m} \Omega z_2 \right ) ~ ; \qquad
b^{\dagger} \equiv \left ({1\over{2 \hbar \Omega}} \right )^{1\over{2}} \left
( {{p_{z_2}}\over{\sqrt{m}}} + i \sqrt{m} \Omega z_2 \right ) ~ ,
\eea
with $[\, a , a^{\dagger} \,] = 1 = [\, b , b^{\dagger} \, ]  , ~~
[\, a , b \,] = 0 = [\, a , b^{\dagger} \, ]$.
Performing
the linear canonical transformation $A \equiv {1\over{\sqrt 2}}
( a + b )$, $B \equiv {1\over{\sqrt 2}} ( a - b )$, with commutation relations $[A,A^{\dagger}] = 1 =  [B,B^{\dagger}], ~~
{[}A,B{]}  =  0\,=\, [A,B^{\dagger}]$,
from Eq.~(\ref{losp2.14}) the Hamiltonian ${\cal H} = {\cal H}_{0} + {\cal H}_{I}$ is obtained~\cite{Celeghini:1992yv}, with
\bea \lab{losp2.21}
{\cal H}_{0} = \hbar \Omega ( A^{\dagger} A - B^{\dagger} B ) = 2 \hbar \Omega {\cal C}~, \qquad
{\cal H}_{I} = i \hbar \Gamma ( A^{\dagger} B^{\dagger} - A B ) = -2 \hbar \Gamma J_{2} ~,
\eea
which is called the {\it fractal Hamiltonian} and where ${\cal C}$  and $J_{2}$ are, respectively, the  Casimir operator and
one of the three generators of the $SU(1,1)$ group  associated
with the system of  quantum oscillators $A$ and $B$. We also have
\be \lab{losp2.25}
[\, {\cal H}_{0} , {\cal H}_{I}\, ] = 0 ~.
\ee
One can see that the eigenvalue of ${\cal H}_{0}$ is the constant (conserved) quantity $\hbar \Omega (n_{A} -
n_{B})$.  The ground state (the vacuum) is $|0 \rangle \equiv | n_{A} = 0 , n_{B} = 0 \rangle$, such that  $A |0\rangle = 0 = B|0\rangle$, and ${\cal H}_{0} |0 \rangle = 0$;
its time evolution is given by
\bea \lab{losp2.25time}
| 0(t) \rangle &=& \exp{\left ( - i t {{{\cal H}}\over{\hbar}}\right )} |0\rangle =
{1\over{\cosh (\Gamma t)}} \exp{\bigl ( \tanh ( \Gamma t ) J_{+} \bigr )}
|0\rangle ~,
\eea
with $J_{+} \equiv A^{\dagger} B^{\dagger}$, namely by a two-mode $SU(1,1)$ generalized coherent state~\cite{Perelomov:1986tf,Celeghini:1992yv} produced by condensation of couples of (entangled) $A$ and $B$ modes: $(AB)^n , \, n = 0,\,1,\, 2\,....\infty$.
At every time
$t$,  $|0(t)\rangle$ has unit norm, $\langle 0(t) | 0(t) \rangle = 1$, however as $t \rightarrow \infty$, the asymptotic state turns out to be orthogonal to the initial state $|0\rangle$:
\be \lab{losp2.25atime}
\lim_{t\to \infty} \langle 0(t) | 0 \rangle = \lim_{t\to \infty} \exp{\bigl ( - \ln
\cosh ( \Gamma t ) \bigr )} \rightarrow 0 ~ ,
\ee
which expresses the  decay (dissipation) of the vacuum under
the time evolution operator ${\cal U} (t) \equiv \exp{\left ( -i t
{{\cal H}_{I} / \hbar}\right )}$ (for large $t$, Eq.~(\ref{losp2.25atime}) gives the ratio $r_0/r(t)$ (cf. Eqs.~(\ref{losp2.2}) and (\ref{losp2.11})) and the limit consistently expresses the (unbounded) growth of $r(t)$).
The state $|0 (t)\rangle$ is known to be a thermal state~\cite{Celeghini:1992yv} and its time evolution
may be written as~\cite{DifettiBook,Celeghini:1992yv}:
\bea \lab{entrop1}
| 0(t) \rangle &=&
\exp{\left ( - {1\over{2}} {\cal S}_{A} (t) \right )} \exp {\left(  A^{\dagger}
B^{\dagger} \right)}|0\rangle
= \exp{\left ( - {1\over{2}} {\cal S}_{B} (t) \right )}
\exp {\left(  A^{\dagger}
B^{\dagger} \right)}|0\rangle ~,
\eea
where ${\cal S}_{A} (t)$ and ${\cal S}_{B} (t)$ have the same formal expressions (with $B$
and $B^{\dagger}$ replacing $A$ and $A^{\dagger}$):
\be \lab{entrop3}
{\cal S}_{A} (t) \equiv -  \Bigl \{ A^{\dagger} A
\ln \sinh^{2} \bigl ( \Gamma t \bigr ) - A
A^{\dagger} \ln \cosh^{2} \bigl ( \Gamma t \bigr ) \Bigr \}
~ ,
\ee
Since $A$'s and $B$'s commute, one simply writes ${\cal S}$ for either ${\cal S}_{A}$ or
${\cal S}_{B}$. ${\cal S}$ in Eq.~(\ref{entrop3}) is recognized to be the
{\sl entropy} for the dissipative system~\cite{Celeghini:1992yv} (see also \cite{Umezawa:1982nv,DifettiBook}).
It is remarkable that the same operator
${\cal S}$ that
controls the
irreversibility of time evolution (breaking of
time-reversal symmetry)  associated with the choice of a privileged
direction in time evolution ({\sl time arrow}), also defines formally the entropy operator.
The breakdown of time-reversal symmetry characteristic of dissipation is clearly manifest in the formation process of fractals; in the case of the logarithmic spiral the breakdown of time-reversal symmetry is associated with the chirality of the spiral: the indirect (right-handed) spiral is the time-reversed, {\it but distinct}, image of the direct (left-handed) spiral (Fig.~\ref{fig2}).

The Hamiltonian ${\cal H}$  then turns out to be the {\it fractal free energy} for the coherent boson condensation process out of which the fractal is formed. By identifying ${\cal H}_0 / \hbar = 2\, \, \Om \, {\cal C}$ with the
``internal energy'' $U$  and $2 \, J_2/ \hbar $ with the entropy $S$, from Eq.~(\ref{losp2.21}) and the defining equation for the temperature $T$ (putting  $k_B = 1$), we have
\be \lab{entrop4} \frac{\partial \, S}{\partial \, U} = \frac{1}{T}
\ee
and obtain $T = \hbar \, \Ga$. Thus, ${\cal H}/ \hbar$ represents the free energy ${\cal F} = U - T \, S$, and the heat contribution in ${\cal F}$ is given by $2 \, \Ga \, J_2$. Consistently, we also have $({\partial \,{\cal F}}/{\partial \, T} )|_{\Om} = - 2 \,J_2 /\hbar$.
It is also interesting that the temperature
$T = \hbar \, \Ga$ is actually proportional to the background zero point
energy: $\hbar \, \Ga \propto \hbar \, \Om /2$~\cite{Blasone:2000ew,Links,DifettiBook}. One can also show that the Planck distribution for the $A$ and $B$ modes is obtained by extremizing the free energy functional ~\cite{Umezawa:1982nv,Celeghini:1992yv,Links,DifettiBook}.

So far for simplicity the
problem has been tackled within the framework of quantum mechanics.  However, it is necessary to stress that the correct  mathematical framework to study quantum dissipation is the quantum field theory (QFT) framework~\cite{Celeghini:1992yv,DifettiBook}, where one considers an infinite number of degrees of freedom. This is  also physically
more realistic, because the realizations of the logarithmic spiral (and in general of fractals) in the many cases it is observed in nature involve an infinite number of elementary degrees of freedom, as it always happens in many-body physics.

The operator ${\cal U}(t)$ written in terms of the $a$ and $b$ operators is~\cite{DifettiBook,CeleghDeMart:1995}
\bea \lab{losp2.26}
{\cal U}(t)
= \exp {\left ( -{\Gamma t \over{ 2}} \Bigl(\bigl(
{a}^2 -
{a}^{\dagger 2}\bigr)
- \bigl(
{b}^2 -
{b}^{\dagger 2}\bigr)
\Bigr) \right )}~,
\eea
and we realize that it is the two mode squeezing generator with squeezing parameter $\zeta = - \Ga \, t$. The $SU(1,1)$ generalized coherent state (\ref{losp2.25time}) is thus a squeezed state.
In the case of the Koch curve, by doubling the degrees of freedom in the operator $q^N$ one obtains the ``doubled" exponential operator
\be \lab{(4.3doub1)}
\biggl( (c^2 - {c^{\dagger 2}})  -
({\tilde c}^2 - {{\tilde c}^{\dagger 2}} ) \biggr) =
- 2 \biggl( C^{\dagger} D^{\dagger} - C D \biggr) ~.
\ee
where ${\tilde c}$ and ${{\tilde c}^\dagger}$ denote the doubled degrees of freedom and $C \equiv {1\over{\sqrt 2}}( c + {\tilde c})$, $D \equiv {1\over{\sqrt 2}} (c - {\tilde c})$.
The second member of this equation is of the same form as the one of the operator appearing in ${\cal H}_{I}$  in Eq.~(\ref{losp2.21}).
The fractal operator is thus proportional to the dissipative time evolution operator $\exp (-it {\cal H}_{I}/\hbar )$ discussed above and the description in terms of generalized $SU(1,1)$ coherent state is recovered also in the case of the Koch curve.

Of course, one can obtain the relation~(\ref{(4.3doub1)}) by starting since the begin from the analysis of the structure of Eq.~(\ref{30}); use of $q = e^{- d\, \theta}$,  with $d$ the fractal dimension, allows us to write the self-similarity equation $q \, \al = 1$ in polar coordinates as
$u = u_0 \, \al \,e^{d\, \theta}$, which is similar to Eq.~(\ref{losp2.1}). Then we may proceed in a similar fashion as we did for the logarithmic spiral, writing the parametric equations for the fractal in the $z$-plane and considering the closed system (cf. Eq.~(\ref{losp2.8})), and so on. This leads us again to the fractal Hamiltonian  (similar to the one in Eq.(\ref{losp2.21})) and fractal free energy and to  $SU(1,1)$ generalized coherent state.

The quantum dynamical scheme here presented seems therefore underlying the {\it morphogenesis} processes which manifest themselves in the macroscopic appearances ({\it forms}) of the fractals.

\section{Noncommutative geometry}

I consider in this section the noncommutative geometry in the plane in the sense analyzed in Ref. \cite{Sivasubramanian:2003xy} (see also \cite{Banerji}), namely considering that a quantum interference phase (of the Aharanov-Bohm type) can be shown~\cite{Sivasubramanian:2003xy,Links,DifettiBook} to
be associated  with the noncommutative plane, a phenomenon which appears to be related to dissipation and squeezing. There is an even deeper level of analysis which relates such noncommutative features to the noncommutativity of the Hopf algebra characterizing $q$-deformed algebraic structures of the type considered in this paper in connection with $q$-deformed coherent states. For brevity, such a level of analysis will not be presented here. The interested reader is referred to Refs.~\cite{CeleghDeMart:1995,DifettiBook,Celeghini:1998a}.

Let  $ (x_1,x_2)  $ represent
the coordinates of a point in a plane and suppose
that they do not commute:
\be
\left[x_1,x_2\right]=iL^2 .
\label{qauntcom1}
\ee
In order to understand the physical meaning of
the geometric length scale $ L  $, one introduces
\be
z = \frac{x_1+ix_2}{L\sqrt{2}}, \quad
\ \ z^* = \frac{x_1-ix_2}{L\sqrt{2}},
\label{quantcom2}
\ee
with non-vanishing commutator $\left[z,z^* \right]=1$.
Then, the Pythagoras'  distance $ \de $ in the noncommutative plane
is
\be
\de^2=x_1^2+x_2^2=L^2(2z^* z+1).
\label{quantcom4}
\ee
From the known properties of the oscillator destruction
and creation
operators
$ z $ and $ z^* $,
it follows that  $\de$ is
quantized in units of the length scale $ L  $
according to
\be
\de_n^2=L^2(2n+ 1)\ \ {\rm where}\ \ n=0,1,2,3,\ldots \ .
\label{quantcom5}
\ee
Thus, in the plane $(x_{1},x_{2})$
we have quantized ``disks'' of
squared radius $\de^{2}_{n}$ and $ L  $ gives the non-zero radius of the ``smallest'' of such disks (see also \cite{Landau}).
In
the path integral formulation of quantum mechanics one
may consider many possible paths with the same initial point and final point. The quantum interference phase $ \vartheta  $ (of the Aharanov-Bohm type)  between two of such paths can be shown~\cite{Sivasubramanian:2003xy,Links,DifettiBook} to be determined by $ L  $ and the enclosed area
$ {\cal A} $, $\vartheta
= \frac{\cal A}{L^2}$.
In the case of the Koch curve, one can show that \cite{NewMat2008,QI}
\be \lab{(4.8z)} {c} \rar {1\over {\sqrt{2}}} \bigl( x + i p_{x}
 \bigr) \equiv {\hat z} ~ ,\quad { c}^\dagger \rar {1\over
{\sqrt{2}}} \bigl( x - i p_{x} \bigr) \equiv {\hat z}^\dagger ~ ,
\quad [{\hat z}, {\hat z}^\dagger ] = 1 ~, \ee
where $p_{x}= -i {d/ dx}$. ${\hat z}^\dagger \leftrightarrow a^\dag$ and  ${\hat z} \leftrightarrow a$
are the usual creation and annihilation operators  in the
configuration representation.
Under the action of the squeezing
transformation, this leads us to obtain~\cite{NewMat2008,QI}
\be \lab{(4.8)} {\hat z} \rar {\hat z}_{q} = {1\over {q \sqrt{2}}} \bigl( x + i
q^{2} p_{x}
 \bigr)  ~ ,\quad {\hat z}^\dagger  \rar {\hat z}^\dagger_{q} = {1\over
{q \sqrt{2}}} \bigl( x - i q^{2} p_{x} \bigr) , \quad [{\hat
z}_{q},{\hat z}^\dagger_{q} ] = 1 ~. \ee
In the ``deformed'' phase space, the
coordinates $x_{1} \equiv x$ and $x_{2} \equiv q^{2}p_{x}$
do not commute:
\be \lab{(4.8)}  [x_{1},x_{2}] = iq^{2}  \ee
(adimensional units and $\hbar = 1$ are used).
Thus the $q$-deformation introduces noncommutative
geometry in the $(x_{1},x_{2})$-plane. We recognize that the $q$-deformation parameter (squeezing) plays the r\^ole of the noncommutative geometric length $L$ and the remarks made above again apply. The noncommutative Pythagora's theorem holds:
\be \lab{(4.9)}  \de^{2} = x_{1}^{2} + x_{2}^{2} = 2 q^{2} \bigl({\hat
z}^\dagger_{q} {\hat z}_{q} + \frac{1}{2} \bigr)~, \ee
and in ${\cal F}$, in the $y \to 0$ limit, (i.e. in the limit $\al \rar
{\it Re} (\al)$,  $\al \equiv x + iy$) the properties of creation and annihilation operators give
\be \lab{(4.10)}  \de^{2}_{n} =  2q^{2} \bigl(n + \frac{1}{2} \bigr)~,
\quad n = 0,1,2,3... ~.\ee
The unit scale in the noncommutative plane is now set by the $q$-deformation parameter.
The quantum interference phase between two alternative paths
in the plane is given by
$\vartheta = {\cal A}/q^2$. The zero point uncertainty relation is also controlled by $ q  $,
$ \Delta x_1 \, \Delta x_2 \ge (q^2/2) $. 

Eq.~(\ref{(4.10)}) actually is the expression of the energy spectrum of the harmonic oscillator and we can  write $\frac{1}{2} \de^{2}_{n} =
q^{2} \bigl(n + 1/2 \bigr) \equiv E_{n}$, $n = 0,1,2,3...$,
where $E_{n}$ might be thought as the ``energy'' associated with the
fractal $n$-stage.

In the case of the two mode description (of the logarithmic spiral as well as the Koch curve),  the interference phase appears as a ``dissipative
interference phase''~\cite{Blasone:1998xt,Sivasubramanian:2003xy}, which provides a relation between dissipation and noncommutative
geometry in the plane.
In the $ (z_1,z_2)$  plane, by using for simplicity the index notation $+ \equiv 1$ and  $- \equiv 2$ and  Eq.~(\ref{losp2.14p}) for the momenta $p_{z_\pm}$, the components  of forward in time and
backward in time velocity $ v_\pm =\dot{z}_\pm $ are given by
\be v_{\pm } = \frac{1}{m} \, ( {p_{z_\mp}} \mp \half \gamma {z_\pm})
\ee
and they do not commute $[v_+,v_-]= - i \,{\ga/ m^2}$.
A canonical set of conjugate position coordinates
$ (\xi_+,\xi_-)$ may be defined by putting $\xi_\pm = \mp (m/\ga) v_\pm $,
so that
\be
\left[\xi_+,\xi_-  \right] =  i \, \frac{1}{\ga} ~. \label{DP2} \ee
Equation (\ref{DP2}) 
characterizes the noncommutative geometry in the plane $(z_+,\, z_-)$.
Thus, since in the present case $L^2 = 1 / \ga$,
the quantum
dissipative phase interference $\vartheta = {\cal A}/L^2 =
{\cal A} \, \ga$ is associated with the two paths ${\cal P}_1$
and ${\cal P}_2$ in the noncommutative plane, provided $z_+ \neq z_-$.

Without adding further details for brevity (see Refs.~~\cite{Celeghini:1998a,CeleghDeMart:1995,Links,DifettiBook})), a final remark is that
the map ${\cal A} \, \rightarrow \,{\cal A}_1
\otimes {\cal A}_2$  which
duplicates the algebra is the Hopf coproduct map ${\cal A} \, \rightarrow \, {\cal A} \otimes \mathbf{1} + \mathbf{1}
\otimes {\cal A}$.  It is known that the Bogoliubov transformations of ``angle"
$\Ga \, t$
are
obtained by convenient combinations of the deformed coproduct
 $\Delta a^{\dag}_q=a^{\dag}_q\otimes q^{1/2} +
q^{-1/2}\otimes a^{\dag}_q$, where  $a^{\dag}_q$ are the creation operators in
the $q$-deformed Hopf algebra~\cite{Celeghini:1998a}. These deformed
coproduct maps are noncommutative and the $q$-deformation parameter is
related
to the coherent condensate content of the state
$|0 (t) \rangle$.  This sheds some light on the physical meaning of the relation between dissipation (which is at the origin of $q$-deformation), noncommutative geometry and the non-trivial topology of paths in the phase space~\cite{Celeghini:1998a,Iorio:1994jk}. It has to be remarked also that, finite difference operators (Eq.~(\ref{(2.12)})) and $q$-algebras which are at the basis of squeezed states~\cite{CeleghDeMart:1995} are  essential tools in the physics of discretized systems. Indeed, it has been recognized~\cite{CeleghDeMart:1995} that $q$-deformed algebras appear, independently on other coexisting mathematical properties, such as analyticity and differentiability, whenever some discreteness in space and/or time plays a r\^ole in the formalism, which is consistent with the case of the fractal self-similarity properties here considered which are related to a limiting process involving discrete structures.

\section{Conclusions}

The relation established in this paper between the self-similarity properties of fractals and coherent states
introduces  dynamical considerations in the study of fractals and
their origin.
Like in the case of extended objects in condensed matter physics~\cite{Umezawa:1982nv,DifettiBook,Bunkov}, also the self-similarity properties of fractals appear to be generated by the dynamical process controlling boson condensation at a microscopic level.
Quantum dissipation, through quantum deformation, or squeezing, of coherent states appears at the root of the observed self-similarity properties at a macroscopic level. To the extent in which fractals are considered under the point of view of self-similarity, as done in this paper, fractals are thus examples of {\it macroscopic quantum systems}, in the specific sense that their macroscopic self-similarity properties cannot be derived without recurring to the underlying quantum dynamics controlled by the  Hamiltonian (Eq.~(\ref{losp2.26})), or, more specifically, by the  fractal free energy.
The fact that the entropy controls the  time-evolution is consistent with the breakdown of time-reversal symmetry
({\it arrow of time}) which is clearly manifest in the formation process of fractals; it is for example related to the chirality in the logarithmic spiral where the indirect (right-handed) spiral is the time-reversed, {\it but distinct}, image of the direct (left-handed) spiral (or vice-versa). These considerations suggest that the quantum dynamics here analyzed is actually at the basis of the {\it morphogenesis} processes responsible of the fractal macroscopic growth.

The links, which also have been discussed, between fractal self-similarity and noncommutative geometry may open interesting
perspectives in many applications where quantum dissipation
cannot be actually neglected, e.g. in condensed matter physics, in quantum
optics and in quantum computing.

I finally recall that the $q$-deformation parameter acts as a label for the
Weyl-Heisenberg representations of the canonical commutation relations. In the limit of infinite degrees of freedom (i.e. in QFT)
different values of $q$  label ``different'', i.e. unitarily
inequivalent representations~\cite{Iorio:1994jk,CeleghDeMart:1995}.
By tuning the value of the
$q$-parameter one thus moves from a given representation to another
one, unitarily inequivalent to the former one. Besides
such a parameter one might also consider phase
parameters and translation parameters characterizing (generalized)
coherent states. By varying these parameters
in a {\it deterministic
iterated function process}, also referred to as {\it multiple
reproduction copy machine} process, the Koch curve, for example, may be transformed into another fractal (e.g.
into Barnsley's fern \cite{Peitgen}). In the frame here presented,
these fractals are described by
unitarily inequivalent representations in the limit of
infinitely many degrees of freedom.
Work in such a direction is still needed and is planned for the future. In the present paper, I have not discussed the measure of lengths in
fractals, the Hausdorff measure, the fractal ``mass'', random fractals,
and other fractal properties.
The discussion has been limited only to self-similarity properties of deterministic fractals.
From a practical point of view, it may be used as a predictive tool: anytime experimental results lead to a straight line of given non-vanishing slope in a log-log plot, one may infer that a specific coherent state dynamics underlies the phenomenon under study. Such a kind of ``theorem'' has been positively confirmed in some applications in neuroscience~\cite{NewMat2008,QI,Freeman}. Other applications are planned to be pursued in the future.

In view of the r\^ole played by coherent states in many applications, ranging from condensed matter physics to quantum optics and molecular biology, elementary particle physics and cosmology, it is also interesting that it is possible to ``reverse'' the arguments  presented in this paper. Indeed, it appears that the conclusion that fractal self-similarity properties may be described in terms of coherent states may be reversed in the statement that coherent states have (fractal) self-similarity properties, namely there is a ``geometry'' characterizing coherent states which  exhibits self-similarity properties.

\section*{Acknowledgements}

This work has been supported financially by INFN and Miur. Useful discussions with Antonio Capolupo are acknowledged.

\appendix

\section{The golden spiral and the Fibonacci progression}

Consider the factor $e^{d \, \theta}$ in Eq.~(\ref{losp2.1}). Suppose that at $\theta = \pi/2$ we have $r/r_0 = e^{d \, (\pi/2)} = \phi$, where $\phi$ denotes the golden ratio, $\phi = (1 + \sqrt{5})/2$.
Then the logarithmic spiral is called the {\it golden spiral}~\cite{Peitgen} and we may put $d_g  \equiv  (\ln \phi)/(\pi/2)$, where the subscript $g$ stays for $golden$. Thus, the polar equation for the golden spiral is $r_{g} (\theta) = r_0 \, e^{d_{g} \, \theta}$.

The radius of the golden spiral grows in geometrical progression of ratio $\phi$ as $\theta$ grows of $\pi/2$: $r_{g}(\theta + n\,\pi/2) =  r_0 \, e^{d_{g} \, (\theta + n\, \pi/2)} =  r_0 \, e^{d_{g} \, \theta} \, \phi^n $ and $r_{g,n} \equiv r_{g} (\theta = n\,\pi/2) = r_0 \, \phi^n $. Here and in the following $n = 0,\, 1,\, 2,\, 3,...$.

A good ``approximate'' construction  of the golden spiral is obtained by using the so called Fibonacci tiling, obtained by drawing in a proper way~\cite{Peitgen} squares whose sides are in the Fibonacci progression, $1, \, 1,\, 2, \, 3, \,5,\, 8,\, 13, ....$ (the Fibonacci generic number in the progression is $F_n  = F_{n-1} + F_{n-2}$, with $~F_0 = 0$; $~F_1 = 1$).
The Fibonacci spiral is then made from quarter-circles tangent to the interior of each square and it does not perfectly overlap with the golden spiral. The reason is that the $F_n /F_{n-1} \rar \phi$ in the $n \rar \infty$ limit, but is {\it not equal} to $\phi$ for given finite $n$ and $n - 1$.

The golden ratio $\phi$ and its ``conjugate'' $\psi = 1 - \phi = - 1 /\phi = (1 - \sqrt{5})/2$ are both solutions of the ``quadratic formula'':
\be \lab{A.1}
x^2 - x - 1 = 0
\ee
and of the recurrence equation $x^n - x^{n-1} - x^{n-2} = 0$,
which, for $n =2$, is the relation (\ref{A.1}).
We thus see that the geometric progression of ratio $\phi$ of the radii $r_{g,n} = r_0 \, \phi^n $  of the golden spiral also satisfy for $n \ge 2$ the recurrence equation
(\ref{A.1}). See ref. \cite{Pashaev} for a recent analysis on $q$-groups and the Fibonacci progression. A final remark is that Eq.~(\ref{A.1}) is the characteristic equation of the differential equation
\be \lab{A.4}
\ddot{r} + \dot{r}  - r = 0~,
\ee
which admits as solution $r (t) = r_0 \, e^{i\,\om \, t} \, e^{+ d\, \theta (t)}$ with
$\om = \pm \, i \, \sqrt{5}/2$ and $\theta = - t/{(2 \, d)} \, + c$, with $c$, $r_0$ and $d$ constants. By setting $c =0$, $r (t) = {r}_0 \, e^{\mp \sqrt{5} \,\, t/2 } \, e^{-  t/2 }$,  i.e. $r_{\phi}(t) = {r}_0 \, e^{- \phi \, t}$ and $r_{\psi} (t) = {r}_0 \, e^{- \psi \, t}$.

\end{document}